\documentclass[aps,prl,reprint,superscriptaddress]{revtex4-1}

\usepackage{graphicx}
\usepackage{dcolumn}
\usepackage{bm}
\usepackage{amsmath}
\usepackage{xfrac}

\begin{document}

\title{Strongly Correlated Quantum Walks in Optical Lattices}

\author{Philipp M. Preiss}
\affiliation{Department of Physics, Harvard University, Cambridge,
Massachusetts, 02138, USA}
\author{Ruichao Ma}
\affiliation{Department of Physics, Harvard University, Cambridge,
Massachusetts, 02138, USA}
\author{M. Eric Tai}
\affiliation{Department of Physics, Harvard University, Cambridge,
Massachusetts, 02138, USA}
\author{Alexander Lukin}
\affiliation{Department of Physics, Harvard University, Cambridge,
Massachusetts, 02138, USA}
\author{Matthew Rispoli}
\affiliation{Department of Physics, Harvard University, Cambridge,
Massachusetts, 02138, USA}

\author{Philip Zupancic}
\altaffiliation{Present address: Institute for Quantum Electronics, ETH Z\"{u}rich,
8093 Z\"{u}rich, Switzerland}
\affiliation{Department of Physics, Harvard University, Cambridge,
Massachusetts, 02138, USA}

\author{Yoav Lahini}
\affiliation{Department of Physics, Massachusetts Institute of Technology, Cambridge,
Massachusetts, 02139, USA}
\author{Rajibul Islam}
\affiliation{Department of Physics, Harvard University, Cambridge,
Massachusetts, 02138, USA}
\author{Markus Greiner}
\email[Email: ]{greiner@physics.harvard.edu}
\affiliation{Department of Physics, Harvard University, Cambridge,
Massachusetts, 02138, USA}


\begin{abstract}
Full control over the dynamics of interacting, indistinguishable quantum particles is an important prerequisite for the experimental study of strongly correlated quantum matter and the implementation of high-fidelity quantum information processing. Here we demonstrate such control over the quantum walk - the quantum mechanical analogue of the classical random walk - in the strong interaction regime. Using interacting bosonic atoms in an optical lattice, we directly observe fundamental effects such as the emergence of correlations in two-particle quantum walks, as well as strongly correlated Bloch oscillations in tilted optical lattices. Our approach can be scaled to larger systems, greatly extending the class of problems accessible via quantum walks.

\end{abstract}


\maketitle

Quantum walks are the quantum-mechanical analogues of the classical random walk process, describing the propagation of quantum particles on periodic potentials \cite{Aharonov1993, Farhi1998}. Unlike classical objects, particles performing a quantum walk can be in a superposition state and take all possible paths through their environment simultaneously, leading to faster propagation and enhanced sensitivity to initial conditions. These properties have generated considerable interest in using quantum walks for the study of position-space quantum dynamics and for quantum information processing \cite{Venegas-Andraca2012}. Two distinct models of quantum walk with similar physical behavior were devised: The discrete time quantum walk \cite{Aharonov1993}, in which the particle propagates in discrete steps determined by a dynamic internal degree of freedom, and the continuous time quantum walk \cite{Farhi1998}, in which the dynamics is described by a time-independent lattice Hamiltonian.

Experimentally, quantum walks have been implemented for photons \cite{Manouchehri2014}, trapped ions \cite{Schmitz2009, Zahringer2010b}, and neutral atoms \cite{Karski2009b, Weitenberg2011, Fukuhara2013a}, among other platforms \cite{Manouchehri2014}. Until recently, most experiments were aimed at observing the quantum walks of a single quantum particle, which are described by classical wave equations.

An enhancement of quantum effects emerges when more than one indistinguishable particle participates in the quantum walk simultaneously. In such cases, quantum correlations can develop as a consequence of Hanbury Brown-Twiss (HBT) interference and quantum statistics, as was investigated theoretically \cite{Bromberg2009, Aaronson2011} and experimentally \cite{Peruzzo2010b, Sansoni2012, Broome2013, Spring2013, Tillmann2013, Crespi2013}. In the absence of interactions or auxiliary feed-forward measurements of the Knill-Laflamme-Milburn type \cite{Knill2001} this problem is believed to lack full quantum complexity, although it can still become intractable by classical computing \cite{Aaronson2011}.

\begin{figure}[t!]
\begin{center}
    \includegraphics[width=0.48\textwidth]{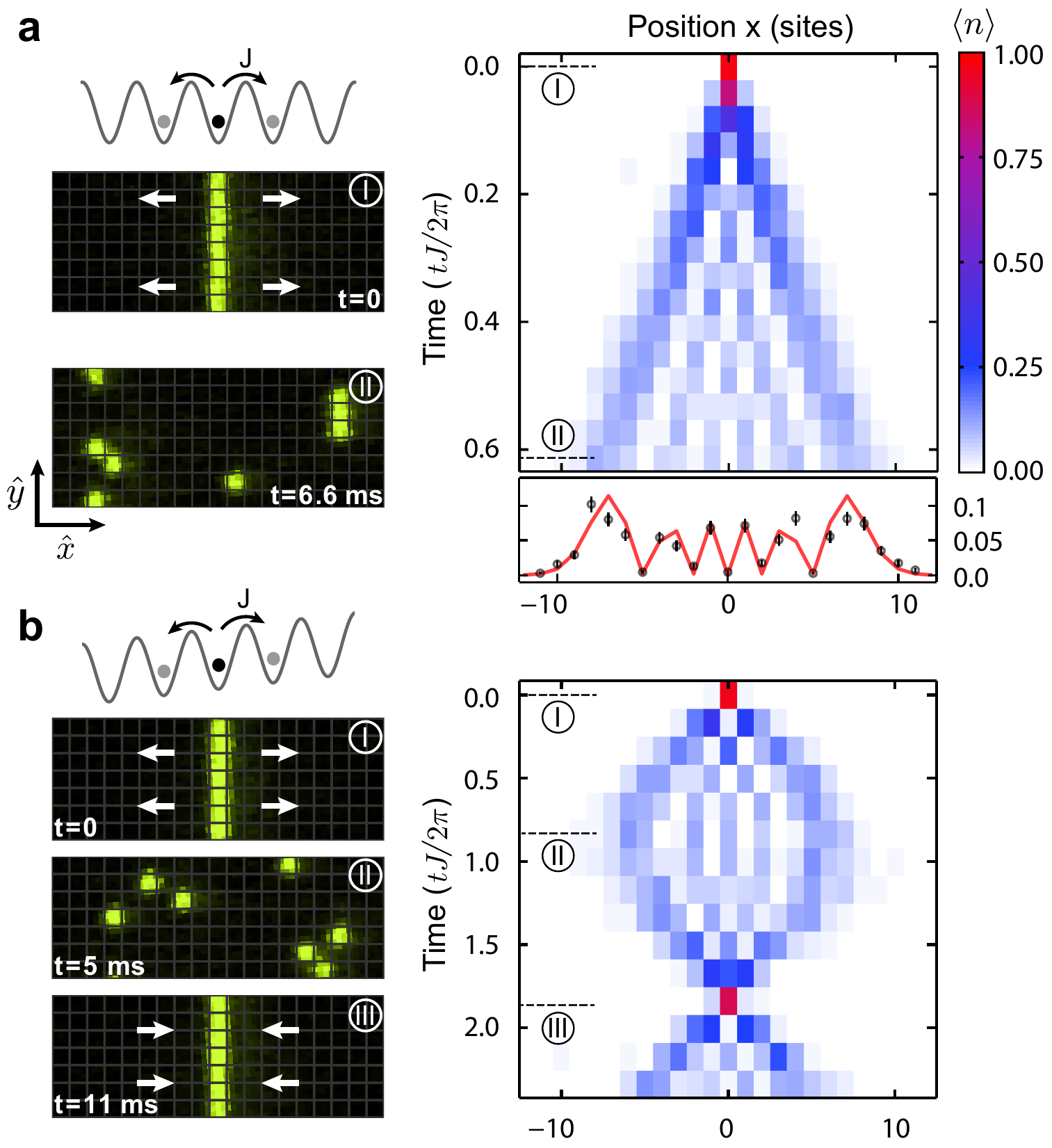}
\end{center}
    \caption{\label{fig:Schematic}Coherent single-particle quantum walks. (a)
    Left: Starting from a localized initial state (I), individual atoms perform independent quantum walks in an optical lattice (II). Right: The single-particle density distribution expands linearly in time, and atoms coherently delocalize over $\sim20$ sites (lower panel shows the averaged density distribution at the end of the quantum walk and a fit to equation \eqref{density} with the tunneling rate $J$ as a free parameter). Error bars: standard error of the mean. (b) In the presence of a gradient, a single
    particle undergoes Bloch oscillations. The atom initially
    delocalizes (II) but maintains excellent coherence and re-converges to its initial position after one
    period (III). Densities are averages over $\sim700$ and $\sim200$ realizations for~a) and~b), respectively.}
\end{figure}

The inclusion of interaction between indistinguishable quantum walkers \cite{Lahini2012b, Ahlbrecht2012} may grant access to a much wider class of computationally hard problems, such as many-body localization and the dynamics of interacting quantum disordered systems \cite{Shepelyansky1994}. Similarly, in the presence of interactions the quantum walk can yield universal and efficient quantum computation \cite{Childs2013b}.

The classical simulation of such correlated quantum dynamics has been achieved with single-particle quantum walks in photonic systems, where effective interactions may be engineered through conditional phase shifts in fiber networks \cite{Schreiber2012} or waveguide arrays \cite{Corrielli2013}. Here, we used bosonic atoms in an optical lattice to directly implement continuous two-particle quantum walks with strong, tunable interactions and direct scalability to larger particle numbers. Our system realizes the fundamental building block of interacting many-body systems with atom-resolved access to the strongly correlated dynamics in a quantum gas microscope \cite{Bakr2009}.

In our experiment, ultracold atoms of bosonic $^{87}$Rb perform quantum walks in decoupled one-dimensional tubes of an optical lattice with spacing $d = 680\,\text{nm}$. The atoms may tunnel in the $x$-direction with amplitude $J$ and experience a repulsive on-site interaction $U$, realizing the Bose-Hubbard Hamiltonian
\begin{equation} \label{Hamiltonian}
    H_{BH}=\sum_{\langle i,j \rangle}-J a_{i}^{\dag} a_{j} + \sum_{i} \frac{U}{2} n_{i}(n_{i}-1)+  \sum_{i}  E\,i\,n_{i}
\end{equation}
Here $a_i^\dag$ and $a_i$ are the bosonic creation and annihilation operators, respectively,  and $n_i  =
a_i^\dag a_i$  gives the atom number on site $i$. The values of $J$ and $U$ are tunable via the depth $V_{x}$ of the optical lattice, specified in units of the recoil energy  $E_r = 2\pi \times \frac{h}{8 m d^2} \approx 2\pi \times 1240\,\text{Hz}$, where $h$ is
Planck's constant and $m$ is the atomic mass of $^{87}$Rb. The energy shift per lattice site $E$ is set by a magnetic field gradient. We measure time in units of inverse tunneling rates, $\tau = t J$, and define the
dimensionless interaction $u = U/J$ and gradient $\Delta = E/J$.

We set the initial motional state of the atoms through an adaptable single-site addressing scheme, enabling the deterministic preparation of a wide range of few-body states: Using a digital micromirror device (DMD) as an amplitude hologram in a Fourier plane, we generate arbitrary diffraction-limited potentials in the plane of the atoms. Starting from a low-entropy two-dimensional Mott insulator with a fixed number of atoms per site, we project a repulsive Hermite-Gauss profile to isolate atoms in selected rows while a short reduction of the optical lattice depth ejects all other atoms from the system (see Methods). For the quantum walk, we prepare one or two rows of atoms along the y-direction of a deep optical lattice with $V_{x}=V_{y} = 45E_r$, (Figure~\ref{fig:Schematic}\,a). The quantum walk is performed at a reduced lattice
depth $V_x$, while the $y$-lattice and the out-of-plane confinement are fixed
at $V_{y} = 45E_{r}$  and $\omega_{z} = 2\pi \times 7.2\,\text{kHz}$, respectively. The atom positions are recorded with single-site resolution using fluorescence imaging in a deep optical lattice \cite{Bakr2009}. Pairs of atoms residing on the same site are lost during imaging because of light-assisted collisions and cannot be detected directly. Using a magnetic field gradient, we separate pairs of atoms along the direction of the quantum walk prior to imaging (see Methods) and obtain the full two-particle correlator $\Gamma_{i,j} = \langle a_{i}^\dag a_{j}^\dag a_i a_j \rangle$ and the density distribution. Only outcomes  with the correct number of atoms per row are included in the data analysis (see Methods).

We first consider quantum walks of individual atoms (Figure \ref{fig:Schematic}\,a). A single particle is initialized at a chosen site in each horizontal tube and propagates in the absence of an external force. For each individual realization the particle is detected on a single lattice site, while the average over many experiments yields the single-particle probability distribution. In contrast to a classical random walk, for which slow, diffusive expansion of the Gaussian density distribution is expected, coherent interference of all single-particle paths leads to ballistic transport with well-defined wavefronts (4). The measured probability density $\rho$ expands linearly in time (Figure~\ref{fig:Schematic}\,a), right panel), in good agreement with the theoretical expectation \cite{Hartmann2004}

\begin{equation} \label{density}
    \rho_{i}(t) = |\mathcal{J}_{i}(2 J t)|^2
\end{equation}
where $\mathcal{J}_{i}$ is a Bessel function of the first kind on lattice site $i$.

If a potential gradient is applied to ultracold atoms in an optical lattice, net transport does not occur due to the absence of dissipation and the separation of the spectrum into discrete bands. Instead, the gradient induces a position-dependent phase shift and causes atoms to undergo Bloch oscillations \cite{Dahan1996}. For a fully coherent single-particle quantum walk with gradient $\Delta$, the atom remains localized to a small volume and undergoes a periodic breathing motion in position space \cite{Hartmann2004, Alberti2009, Haller2010, Genske2013} with a maximal half width $L_{B}=4/\Delta$ and temporal period $T_{B}=2\pi/\Delta$ in units of the inverse
tunneling. Figure \ref{fig:Schematic}\,b) shows a single-particle quantum walk with $\Delta = 0.56 $, resulting in Bloch oscillations over $\sim14$~lattice sites. We observe a high quality revival after one Bloch period and detect the particle back at the origin with a probability of up to 0.96(3) at $\tau=T_B$ in individual tubes. The average over six adjacent rows in Figure \ref{fig:Schematic}\,b) displays a revival probability of 0.88(2), limited by the temporal resolution of the measurements and inhomogeneous broadening across different rows. The fidelity,
\begin{equation} \label{fidelity}
    F(t) = \sum_x \sqrt{p_x(t) q_x(t)}
\end{equation}
for the measured and expected probability distributions $p_x(t)$ and $q_x(t)$, averaged over $\sim$ 1.5 Bloch oscillations is $98.1(1)\%$, indicating that a high level of coherence is maintained while the particle delocalizes over $\sim 10\,\mu$m in the optical lattice.

 \begin{figure*}[t!]
    \includegraphics[width=\textwidth]{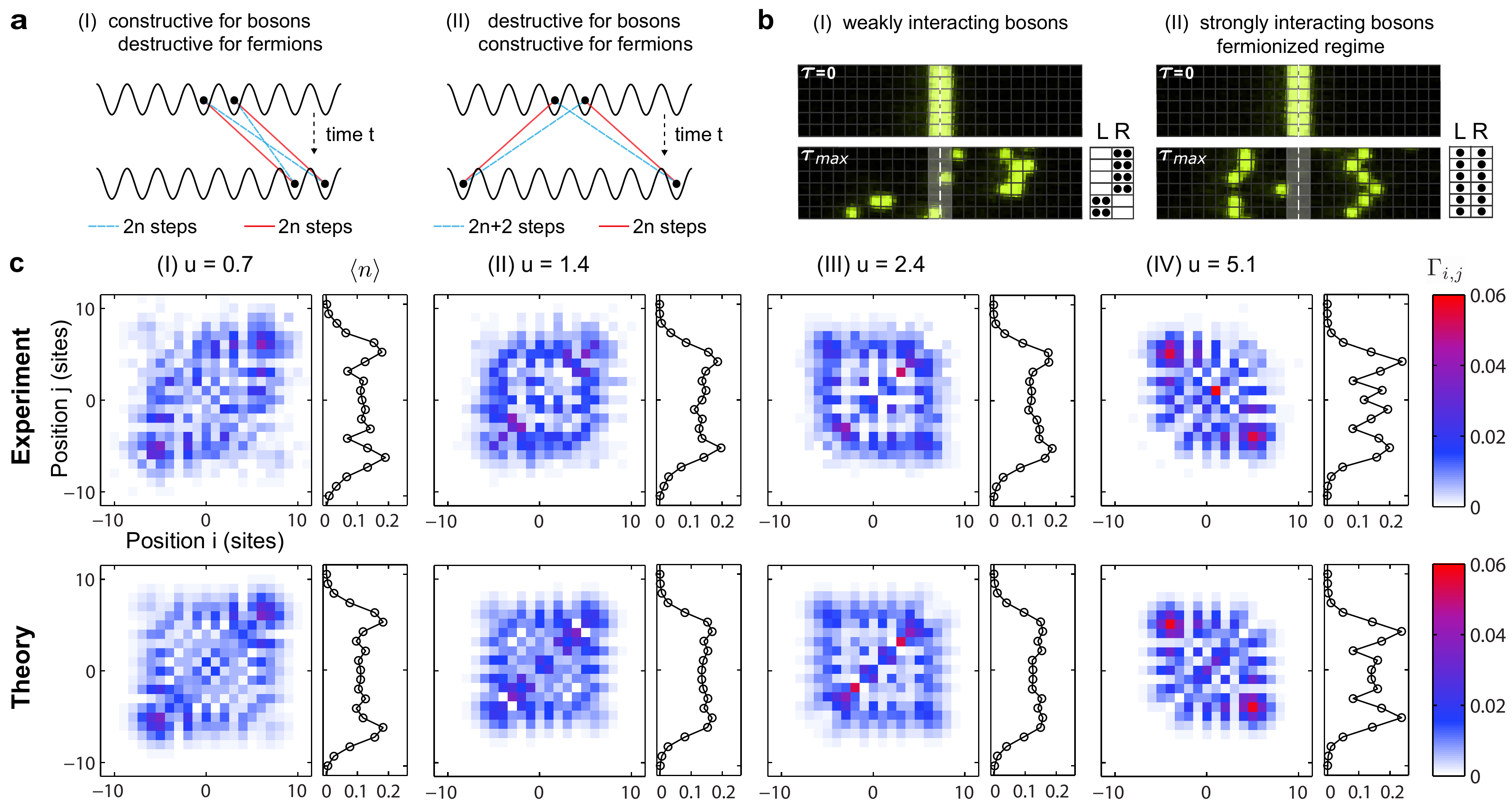}
    \caption{Hanbury Brown-Twiss interference and fermionization. (a) Processes connecting the initial and
    final two-particle states interfere coherently. Each tunneling step contributes a phase $i$. For non-interacting bosons, processes of the same length add constructively~(I), while processes differing in length by two steps interfere destructively (II). (b) Weakly interacting bosons display strong bunching~(I). Strong, repulsive on-site interactions cause bosons in one dimension to fermionize and develop long-range anti-correlations~(II). (c) Measured correlator $\Gamma_{i,j} $ at time $\tau_{max} \approx 2\pi \times
    0.5$, averaged over $\sim 3200$ realizations. The interactions are tuned from weak ($u < 1$) to strong ($u \gg 1$) by choosing
    $V_{x}= 1\,E_r, 2.5\,E_r, 4\,E_r$, and $6.5\,E_r$.}
    \label{fig:QW_correlations}
 \end{figure*}

If two particles undergo a quantum walk simultaneously, the dynamics are
sensitive to the underlying particle statistics due to HBT interference \cite{Peruzzo2010b,Jeltes2007}. All two-particle processes in the system add coherently,
leading to quantum correlations between the particles, shown in Figure~\ref{fig:QW_correlations}\,a): For bosons, the processes
bringing both particles into close proximity of each other add constructively,
leading to bosonic bunching, as observed in tunnel coupled optical tweezers \cite{Kaufman2014}, expanding atomic clouds \cite{Folling2005,Jeltes2007} and photonic implementations of quantum walks \cite{Peruzzo2010b,Sansoni2012}.

In our experiment, the bunching of free bosonic atoms is apparent in single shot
images of quantum walks with two particles starting from adjacent sites in the
state $a_0^\dag a_1^\dag |0\rangle$. For weak interactions, the
two atoms are very likely to be detected close to each other because of HBT
interference, as shown in raw images in Figure \ref{fig:QW_correlations}\,b). We characterize the degree of bunching using the density-density correlator $\Gamma_{i,j}$  in
Figure~\ref{fig:QW_correlations}\,c), measured at time $\tau_{max} \approx 2\pi \times
0.5$. Panel~I shows the two-particle correlator for a quantum walk with weak interactions ($u=0.7$). Sharp features are caused by quantum interference and demonstrate the good coherence of the two-particle dynamics. The concentration of probability on and near the diagonal of the correlator $\Gamma_{i,j}$ indicates HBT interference of nearly free bosonic particles.  
    
We use the sensitivity of the quantum walk to quantum statistics to probe the ``fermionization" of bosonic particles caused by repulsive interactions in one-dimensional systems. When such interactions are strong, double occupancies are suppressed by the large energy cost $U$, which takes the role of an effective Pauli exclusion principle for bosonic particles. In the limiting case of infinite,``hard-core" repulsive interactions, one-dimensional bosonic systems ``fermionize" and show densities and spatial correlations that are identical to those of non-interacting spinless fermions \cite{Cazalilla2011}. This behavior has been observed in equilibrium in the pair-correlations and momentum distributions of large one-dimensional Bose-Einstein Condensates \cite{Paredes2004, Kinoshita2004}. These systems are characterized by the dimensionless ratio of interaction to kinetic energy $\gamma$, and the fermionized Tonks-Girardeau regime is entered when $\gamma$ is large. For Bose-Hubbard systems below unity filling, such as ours, the corresponding parameter is the ratio $u=U/J$.

\begin{figure*}[ht]
    \includegraphics[width=\textwidth]{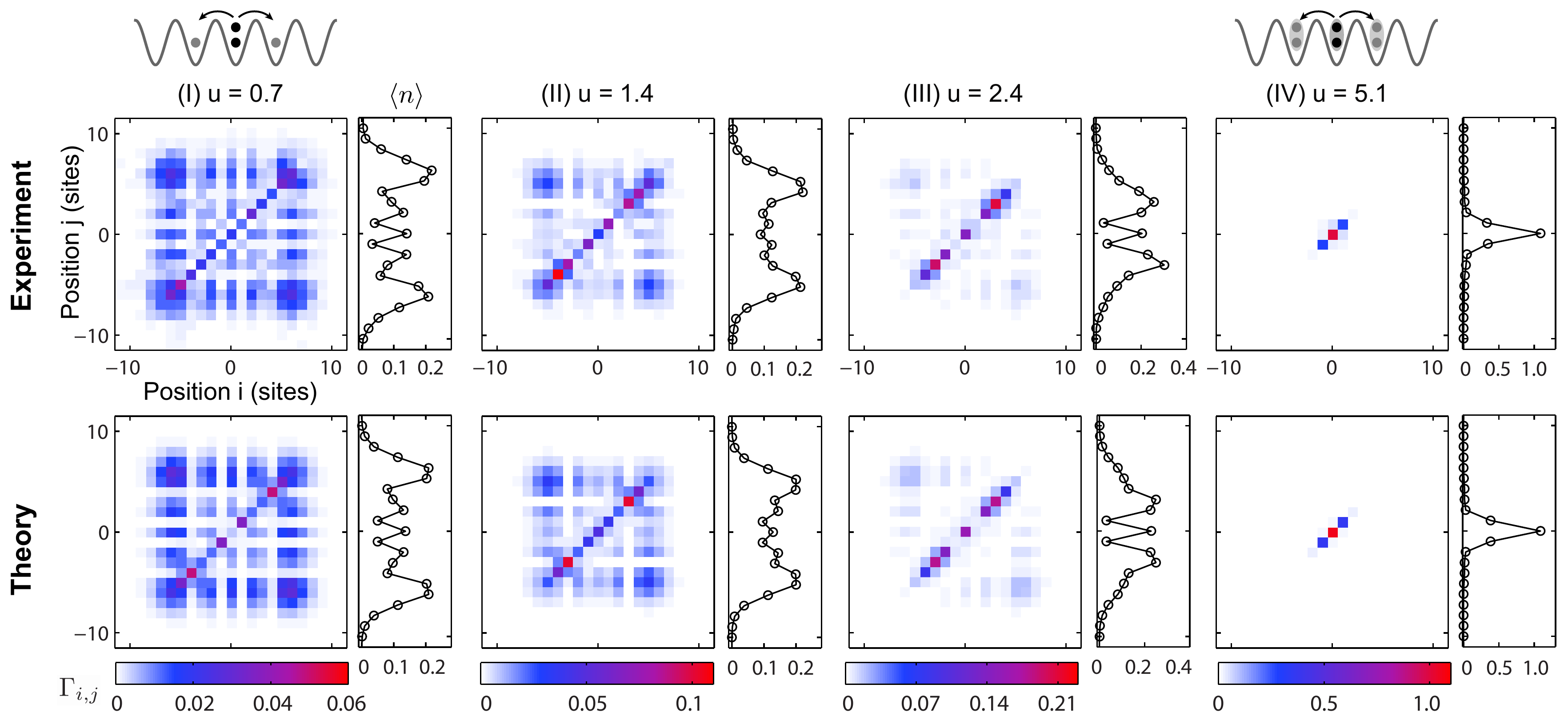}
    \caption{\label{fig:RPB}Formation of repulsively bound pairs.
    Two-particle correlations at $\tau_{max}\approx 2\pi\times 0.5$ for two
    particles starting on site~0 in state $\frac{1}{\sqrt{2}}a_{0}^\dag a_{0}^\dag
|0\rangle$. For weak interactions ($u=0.7$), the atoms perform independent single-particle quantum walks. As the interaction
    strength is increased, repulsively bound pairs form and undergo an effective
    single-particle quantum walk along the diagonal of the two-particle
    correlator. Experimental parameters are identical to those in Figure
    \ref{fig:QW_correlations}.}
    \end{figure*}

We study the process of fermionization in the fundamental unit of two interacting particles by
repeating the quantum walk from initial state $a_0^\dag a_1^\dag |0\rangle$ at increasing
interaction strengths \cite{Lahini2012b}. Figure~\ref{fig:QW_correlations}\,c) shows $\Gamma_{i,j}$ for several values of
$u$. At intermediate values of the interaction $u=1.4$ and $u=2.4$, the
correlation distribution is relatively uniform, as repulsive interactions
compete with HBT interference. For the strongest interaction strength $u=5.1$,
most of the weight is concentrated on the anti-diagonal of $\Gamma_{i,j}$,
corresponding to pronounced anti-bunching. The anti-correlations are strong
enough to be visible in raw images of the quantum walk as in panel~II of
Figure \ref{fig:QW_correlations}\,b), and $\Gamma_{i,j}$ is almost identical to the
expected outcome for non-interacting fermions. Note that although the correlations
change dramatically with increasing interaction, the densities remain largely
unchanged. At all interaction strengths, the observed densities and
correlations are in excellent agreement with a numerical integration of the
Schr\"{o}dinger equation with Hamiltonian (1) (see Methods).
Interactions in two-particle scattering events, which we observe on a lattice, take on a central role in closely related models that may be solved via the Bethe ansatz \cite{Cazalilla2011}, such as Heisenberg spin chains \cite{Fukuhara2013a} and bosonic continuum systems \cite{Kinoshita2004}: Within integrable models, scattering between arbitrary numbers of particles may be decomposed into two-particle scattering events, and the phase shift acquired in such processes determines the microscopic and thermodynamic properties of the system.

The precise control over the initial state in our system enables the study of
strongly interacting bosons in scenarios not described by fermionization,
such as the quantum walks of two atoms prepared in the same state.
Figure~\ref{fig:RPB} shows the correlations and densities for
the initial state $\frac{1}{\sqrt{2}}a_{0}^\dag a_{0}^\dag
|0\rangle$. Because both atoms originate from the same site, HBT interference
terms are not present. In the weakly interacting regime ($u=0.7$),
both particles undergo independent free dynamics and the correlator is the
direct product of the single-particle densities. As the interaction increases,
separation of the individual atoms onto different lattice sites becomes energetically forbidden. 
The two atoms preferentially propagate through the lattice together as reflected in increasing weights on the diagonal of the correlation matrix. For the strongest interactions, the particles form a repulsively bound pair with effective single-particle behavior \cite{Winkler2006b}. The two-particle dynamics may be described as a quantum walk of the bound
pair \cite{Lahini2012b,Ahlbrecht2012} at a decreased tunneling rate $J_{\text{pair}}$ , which reduces to the second-order tunneling \cite{Folling2007} $J_{\text{pair}}=\frac{2J^2}{U} \ll J$ for large values of $u$.

\begin{figure}
\begin{center}
    \includegraphics[width=0.5\textwidth]{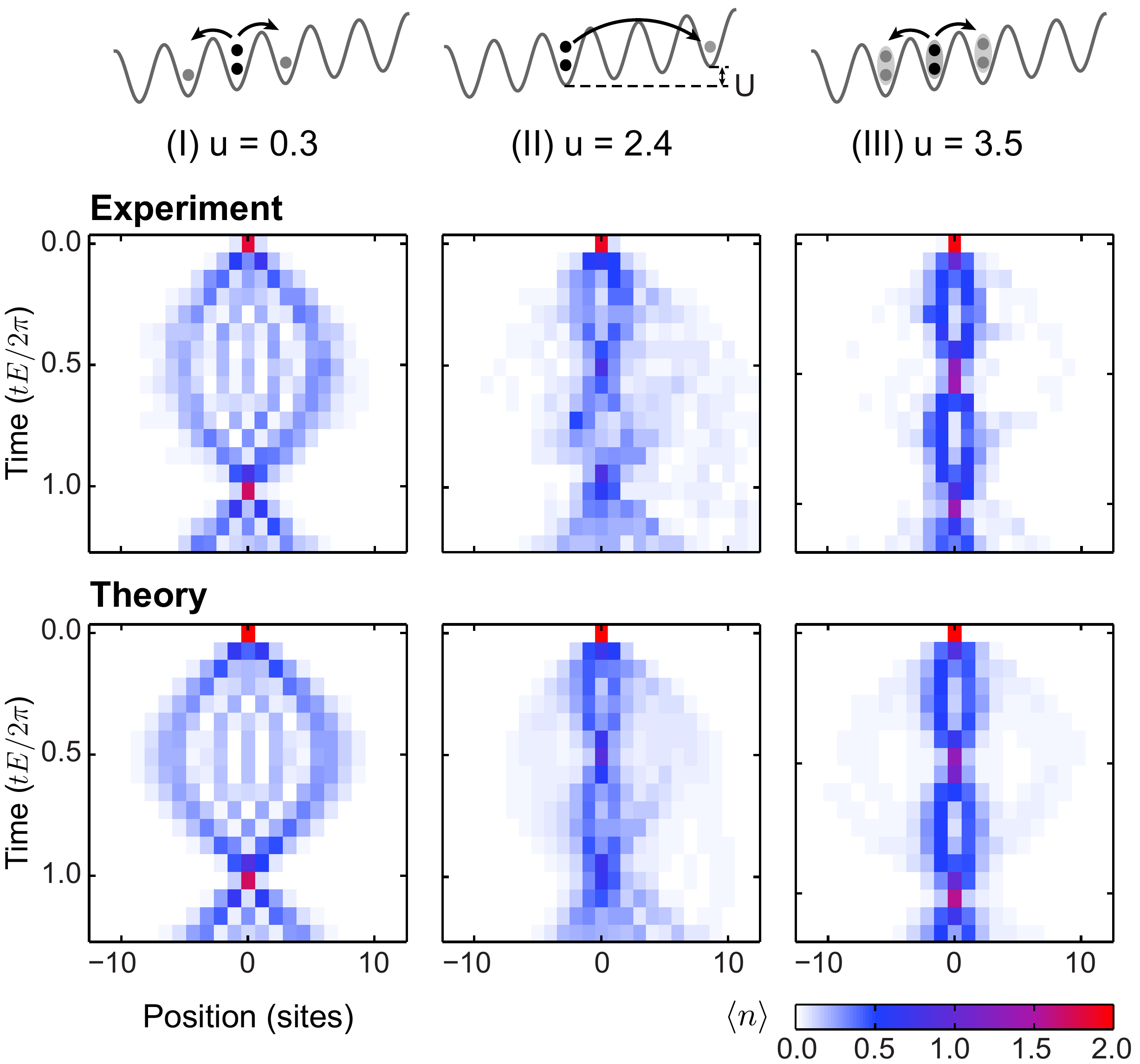}
\end{center}
    \caption{\label{fig:bloch_osc}Bloch oscillations of repulsively bound pairs. In the weakly interacting regime (I), two particles initialized on the same site undergo clean, independent Bloch oscillations. Increasing the interaction strength (II) leads to complex dynamics: Pairs of atoms remain bound near the origin or they separate, breaking left-right symmetry via long-range tunneling.For the largest interactions (III), repulsively bound pairs perform coherent, frequency-doubled Bloch oscillations. Densities are averages over $\sim220$ independent quantum walks.}
\end{figure}

The formation of repulsively bound pairs and their coherent dynamics
can be observed in two-particle Bloch oscillations. We focus on the dynamics
of two particles initially prepared on the same site with a
gradient $\Delta  \approx 0.5$ (\ref{fig:bloch_osc}). In the
weakly interacting regime ($u=0.3$), both particles undergo
symmetric Bloch oscillations as in the single-particle case, and we observe a
high-quality revival after one Bloch period. For intermediate interactions
($u=2.4$), the density evolution is complex: In this regime
where $J$, $U$, and $E$ are similar in magnitude, states both with and without
double occupancy are energetically allowed and contribute to the dynamics.
The skew to the right against the applied force is due to resonant long-range tunneling of single particles over several sites \cite{Khomeriki2010a,Meinert2014} and agrees with numerical simulation.

When the interactions are sufficiently strong ($u=3.5$), the pairs of atoms
are tightly bound by the repulsive interaction and behave like a single
composite particle. However, the effective gradient has doubled with respect to the single-particle case, and the pairs perform Bloch oscillations at twice the fundamental frequency and reduced spatial amplitude. The frequency-doubling of Bloch oscillations was predicted for electron systems \cite{Dias2007} and cold atoms \cite{Ahlbrecht2012, Khomeriki2010a} and has
recently been simulated with photons in a waveguide array \cite{Corrielli2013}. Throughout the breathing motion, the repulsively bound pairs themselves undergo coherent dynamics and delocalize without unbinding. The clean revival after half a Bloch period directly demonstrates the spatial entanglement of atom pairs during the oscillation. 

Quantum walks of ultracold atoms in optical lattices offer an ideal starting point for the ``bottom up" study of many-body quantum dynamics. The present two-particle implementation provides intuitive access to essential features of many-body systems, such as localization caused by interactions or fermionization of bosons. Such microscopic features, when scaled to larger system sizes, manifest in emergent phenomena -- for example, quantum phase transitions, quasi-particles, or superfluidity -- as observed in other cold atom experiments. The particle-by-particle assembly of interacting systems may give access to the crossover from few- to many-body physics and may reveal the microscopic details of disordered quantum systems \cite{Shepelyansky1994} and many-body quench dynamics \cite{Ronzheimer2013}.

We thank S. Aaronson, M. Endres, and M. Knap for helpful discussions. Supported by grants from NSF through the Center for Ultracold Atoms, the Army Research Office with funding from the DARPA OLE program and a MURI program, an Air Force Office of Scientific Research MURI program, the Gordon and Betty Moore Foundation's EPiQS Initiative, the U.S. Department of Defense through the NDSEG program (M.E.T.), a NSF Graduate Research Fellowship (M.R.), and the Pappalardo Fellowship in Physics (Y.L.)

\bibliography{qw_preiss_arxiv}

\clearpage

\subsection*{Methods}
\subsubsection*{State Initialization.}
We start with a two-dimensional Mott insulator in
the $|F,m_{F}\rangle=|1,-1\rangle$ hyperfine state with one or two atoms per site, prepared in a deep lattice ($V_{x}=V_{y}=45\,E_{r}$). Using a digital micromirror device (DMD) in a Fourier plane as an amplitude hologram, we  generate arbitrary optical potentials with single-site resolution. We superimpose a blue-detuned beam ($\lambda=760$\,nm) with a Hermite-Gauss profile along $x$ (waist 700\,nm) and a flattop profile along $y$ (half length 5\,$\mu$m), with a typical peak depth of $25\,E_{r}$. Subsequently, we switch off the $x$-lattice in the presence of a large anti-confining beam for 40\,ms. Only atoms in rows coinciding with the nodes of the Hermite-Gauss beam are retained, while all other atoms are expelled from the system before the $x$-lattice is ramped back on. We thus deterministically prepare one or two rows of atoms along $y$ (length $\approx$10 sites), with a typical single-site loading fidelity of $98\%$. In each experimental run, we realize $6-8$ independent one-dimensional quantum walks in decoupled adjacent tubes.

\subsubsection*{Data Analysis.}

Light-assisted collisions during imaging lead to the pairwise loss of atoms (``parity projection"), preventing us from directly detecting pairs of atoms on the same site \cite{Bakr2009}. Before data analysis, we post-select outcomes for which the atom distribution after the quantum walk is consistent with this parity projection. For single-particle quantum walks, we keep only one-dimensional tubes with exactly one atom. For two-particle quantum walks, we retain outcomes with either two or zero atoms, and assume the latter always corresponds to atom loss due to parity projection only.

The two-particle correlator is obtained from a histogram of the imaged atom positions. Because of parity projection, we cannot measure the diagonal elements $\Gamma_{i,i}$ directly. To circumvent this, we split pairs of atoms prior to imaging for half of the data set: Doubly occupied sites are converted into two atoms on neighboring sites along $x$ and vice versa ($(2,0) \Leftrightarrow (1,1)$) with high fidelity by ramping a magnetic field gradient from $\Delta \approx 0.5 u$ to $\Delta \approx 2 u$ at a reduced lattice depth of $V_{x}=16\,E_{r}$ in 200\,ms \cite{Simon2011}. This process amounts to swapping the diagonal and first off-diagonal of the two-particle correlator prior to imaging. Therefore the on-site correlation $\Gamma_{i,i}$ is obtained from the first off-diagonal elements of the histogram with the density mapping, while $\Gamma_{i,i+1}$ is determined directly from the histogram without the density mapping. To get the full correlator, we combine the two histograms weighted by the number of post-selected realizations in each half of the data set.

The aforementioned assumption in post-selection ensures the proper normalization of the histograms and is verified by comparing the far off-diagonal elements in the two weighted histograms, which are not affected by the density mapping and typically differ by less than 3\%.

For two particles, the density distribution $\left\langle n_i\right\rangle$ is then obtained by summing the correlator along one axis: $\left\langle n_i\right\rangle = \sum_j{\Gamma_{i,j}}$.

\subsubsection*{Bose-Hubbard Parameters.}

We initially calibrate lattice depths using Kapitza-Dirac scattering with an uncertainty of $10\%$. Single-particle Bloch oscillations serve as our most sensitive probe of the tunneling $J$ with a typical uncertainty of $5\%$, in agreement with a band structure calculation. The interaction $U$ is measured at 14\,$E_r$ with photon-assisted tunneling in a tilted lattice \cite{Ma2011}, and extrapolated to other lattice depths using a numerical calculation.

All theory plots are obtained from a direct numerical solution of the Schr\"odinger equation with  Hamiltonian (1) in the Fock space of two particles on 23~lattice sites. The values of  $U$ and $t_{max}$ are fixed, while $J$ (and $E$ in the case of Bloch oscillations) are left as free parameters to minimize the rms error between measured and calculated densities. 

The minimization is performed simultaneously on data sets from Figures 2 and 3, except for panel~IV. Parameter values for all data sets are listed in Table S1. The fitted values $J_{fit}$ are generally in good agreement with the measurements from single-particle dynamics. At low lattice depths of $1-3\,E_r$, next-nearest-neighbor hopping is significant, resulting in dynamics up to $20\%$ faster than expected from Hamiltonian (1). For the deep lattice at $6.5\,E_r$, residual gradients of $\sim 20$\,Hz/site affect the dynamics, leading to a slower quantum walk than for $\Delta=0$ and to the strong peak near the origin in panel~IV of Figure 2~c).

\begin{table*}[h]
\small
\begin{tabular}{l*{6}{c}}
\hline
\\
Data Set            &  $V_x$ [$E_r$] 	& $J_{\text{sp}}/(2\pi$) [Hz] 	&  $U/(2\pi$)
[Hz]	&   $J_{\text{fit}}/(2\pi$) [Hz]  	& $ E_{\text{fit}}/(2\pi$) [Hz] &  $t_{max}$[ms] \\
\hline
\\
Fig. 1 a 					&4.5 	& 97(6)		&  -		& 107	&  -	& 6.6  \\
Fig. 1 b 					& 2.5	&160(9) 			&  -		& 166 	& 93 	& 14  \\
Fig. 2 \& 3 (I) 	&1  	& 227(12)			& 161		& 274 	&  -	& 2.1  \\
Fig. 2 \& 3 (II) 	& 2.5 	&  160(9) 			&216		& 168 &  -	& 3.0  \\
Fig. 2 \& 3 (III) 	& 4 	& 108(4) 			& 255		& 109 &  -	& 5.0  \\
Fig. 2 (IV)			& 6.5	&  59(3)  	& 299 		& 42	&  -	& 10.9  \\
Fig. 3 (IV) 					& 6.5 	&  59(3)		& 299 		& 34 	&  -	& 10.9  \\
Fig. 4  (I)				& 2.5 	& 160(9) 			&53 		&173  	& 97 	& 12.8  \\
Fig. 4  (II)				& 4  	&108(4) 			&255 		&101 	& 54 	& 22.8  \\
Fig. 4  (III)				& 5 	& 80(6) 			& 279		& 81	& 34 	& 34  \\
\hline
\end{tabular}
\caption{\label{tab:Simulations} Bose-Hubbard parameters used for theory plots. $V_x$ are approximate lattice depths. $J_{\text{sp}}$ is the nearest-neighbor tunneling obtained from single-particle Bloch oscillations or directly from a band structure calculation. Typical errors on $U$ are $3\%$ from the uncertainty in the calibration. $J_{\text{fit}}$  and $E_{\text{fit}}$ are the results from fitting density distributions.}
  
\end{table*}

\end{document}